\documentclass[aps,pra,twocolumn,superscriptaddress]{revtex4-2}
\usepackage[utf8]{inputenc}
\usepackage{soul}
\usepackage{textcomp}
\usepackage[english]{babel}
\usepackage{graphicx}
\usepackage{dcolumn}
\usepackage{bm}
\usepackage{amsmath}
\usepackage{verbatim}     
\usepackage[dvipsnames]{xcolor}       
\usepackage[normalem]{ulem}   
\usepackage{bbold}
\usepackage{mathrsfs}
\usepackage[mathscr]{eucal}
\usepackage{physics}
\usepackage{algorithmic}

\usepackage{hyperref}
\hypersetup{
   pdfpagemode=None,
   pdfstartpage=1,
   pdfmenubar=true,
   pdftoolbar=true,
   colorlinks = true,
   linkcolor=blue,
   citecolor=blue,
   urlcolor=blue,
   bookmarksopen=false
 } 

 \usepackage{booktabs}
 \usepackage{float}

\definecolor{mirko}{rgb}{0.6,0.1,0.05}
\newcommand{\ma}[1]{{\color{mirko}#1}}

\begin{document}
\title{Engineering qubit dynamics in open systems with photonic synthetic lattices}

\author{Francesco Di Colandrea}
\email{francesco.dicolandrea@unina.it}
\affiliation{Nexus for Quantum Technologies, University of Ottawa, K1N 5N6, Ottawa, ON, Canada}
\affiliation{Dipartimento di Fisica, Universit\`{a} degli Studi di Napoli Federico II, Complesso Universitario di Monte Sant'Angelo, Via Cintia, 80126 Napoli, Italy}

\author{Tareq Jaouni}
\affiliation{Nexus for Quantum Technologies, University of Ottawa, K1N 5N6, Ottawa, ON, Canada}

\author{John Grace}
\affiliation{Nexus for Quantum Technologies, University of Ottawa, K1N 5N6, Ottawa, ON, Canada}
\affiliation{McGill University, H3A 0G4, Montreal, QC, Canada}

\author{Dilip Paneru}
\affiliation{Nexus for Quantum Technologies, University of Ottawa, K1N 5N6, Ottawa, ON, Canada}

\author{Mirko Arienzo}
\affiliation{Institute for Quantum-Inspired and Quantum Optimization, Hamburg University of Technology, 21079 Hamburg, Germany}

\author{Alessio D'Errico}
\affiliation{Nexus for Quantum Technologies, University of Ottawa, K1N 5N6, Ottawa, ON, Canada}
\affiliation{National Research Council of Canada, 100 Sussex Drive, K1A 0R6, Ottawa, ON, Canada}

\author{Ebrahim Karimi}
\affiliation{Nexus for Quantum Technologies, University of Ottawa, K1N 5N6, Ottawa, ON, Canada}
\affiliation{National Research Council of Canada, 100 Sussex Drive, K1A 0R6, Ottawa, ON, Canada}
\affiliation{Institute for Quantum Studies, Chapman University, Orange, California 92866, USA}


\begin{abstract} 
The evolution of a quantum system interacting with an environment can be described as a unitary process acting on both the system and the environment. In this framework, the system's evolution can be predicted by tracing out the environmental degrees of freedom. Here, we establish a precise mapping between the global unitary dynamics and the quantum operation involving the system, wherein the system is a single qubit, and the environment is modeled as a discrete lattice space. This approach enables the implementation of arbitrary noise operations on single-polarization qubits using a minimal set of three liquid-crystal metasurfaces, whose transverse distribution of the optic axes can be patterned to reproduce the target process. We experimentally validate this method by simulating common noise processes, such as phase errors and depolarization.

\end{abstract}

\maketitle


\section{Introduction}
The majority of experiments in quantum mechanics require, to some extent, to account for the influence of external effects over which the experimenter has limited knowledge or control. At the same time, the interaction of a system of one or few quantum bits with an environment involves exchanges of energy, work, and information, which are at the basis of quantum thermodynamics~\cite{kosloff2013quantum, vinjanampathy2016quantum,deffner2019quantum}. Engineering qubit-environment interactions is thus a pivotal task for devising energy transfer processes, such as quantum heat engines~\cite{scully2001extracting, quan2007quantum,scully2011quantum,fialko2012isolated,peterson2019experimental}, quantum batteries~\cite{alicki2013entanglement, binder2015quantacell,campaioli2024colloquium, catalano2024frustrating}, and quantum refrigerators~\cite{kosloff2000quantum,levy2012quantum,brunner2014entanglement,correa2014quantum}. 

Experimental setups where such interaction can be partially controlled have recently been demonstrated. For instance, the environment can be cooled down using active feedback control loops on a qubit~\cite{spiecker2023two}, thus allowing for reducing decoherence effects in quantum devices~\cite{bluhm2010enhancing,broadway2018quantum}. 

In recent years, it has also become common to describe open systems in terms of non-Hermitian processes, wherein the system evolution is dictated by a Hamiltonian operator that admits a complex spectrum, accounting for the possibility of gains and losses in subsets of modes~\cite{moiseyev2011non,ashida2020non,hatano1997vortex}. Non-Hermitian dynamics have demonstrated a wide variety of applications in quantum physics~\cite{hatano1996localization,hamazaki2019non,lee2020many,kawabata2022many,bu2024chiral} and photonics~\cite{feng2017non,li2023exceptional,harari2018topological}, in particular when exploiting phenomena related to the topological properties of the Hamiltonian~\cite{gong2018topological,bergholtz2021exceptional,okuma2023non}. The connection between the description of open systems as non-Hermitian processes and via Master equations has been established~\cite{minganti2019quantum}, proving that it is generally possible to reduce the problem to a set of Kraus operators acting on the principal system~\cite{nielsen2010quantum}.

Engineering system-environment interactions can also be of interest for investigating fundamental questions, for instance, in the context of emerging developments in understanding the interplay between geometric phases and quantum back action~\cite{cho2019emergence}, wherein the variation of the geometrical phase has been demonstrated to be connected with the system-environment coupling strength in a topological fashion~\cite{ferrer2023topological}.

Different schemes have been proposed to engineer open quantum systems, but they still lack full generality~\cite{10.1063/1.4977023,Nokkala_2018,PhysRevLett.127.020504,PRXQuantum.4.040310}. In this work, we demonstrate a step forward in this direction. We realize a platform where arbitrary sets of Kraus operators acting on a single qubit can be implemented, thus allowing for the simulation of any noise operator. Our strategy is based on coupling the qubit with an environment modeled as an infinite, discrete set of modes, visualized as a one-dimensional (1D) lattice space. By leveraging translational invariance, we derive an effective relation between the global (system-environment) unitary and the target set of Kraus operators. Once such a unitary is identified, it can be implemented in a compact photonic circuit consisting of a stack of three patterned waveplates~\cite{DiColandrea2023}. 
In so doing, we can implement arbitrary operations on light polarization, encoding the qubit, by coupling it with the transverse spatial degree of freedom. This method is validated in a proof-of-principle experiment where we simulate common noise processes with different strengths. Our results demonstrate that this approach could also be employed to benchmark quantum thermal machines' functionality and test the fundamental properties of open quantum systems.

\section{Results}
\begin{figure}
    \centering
\includegraphics[width=\columnwidth]{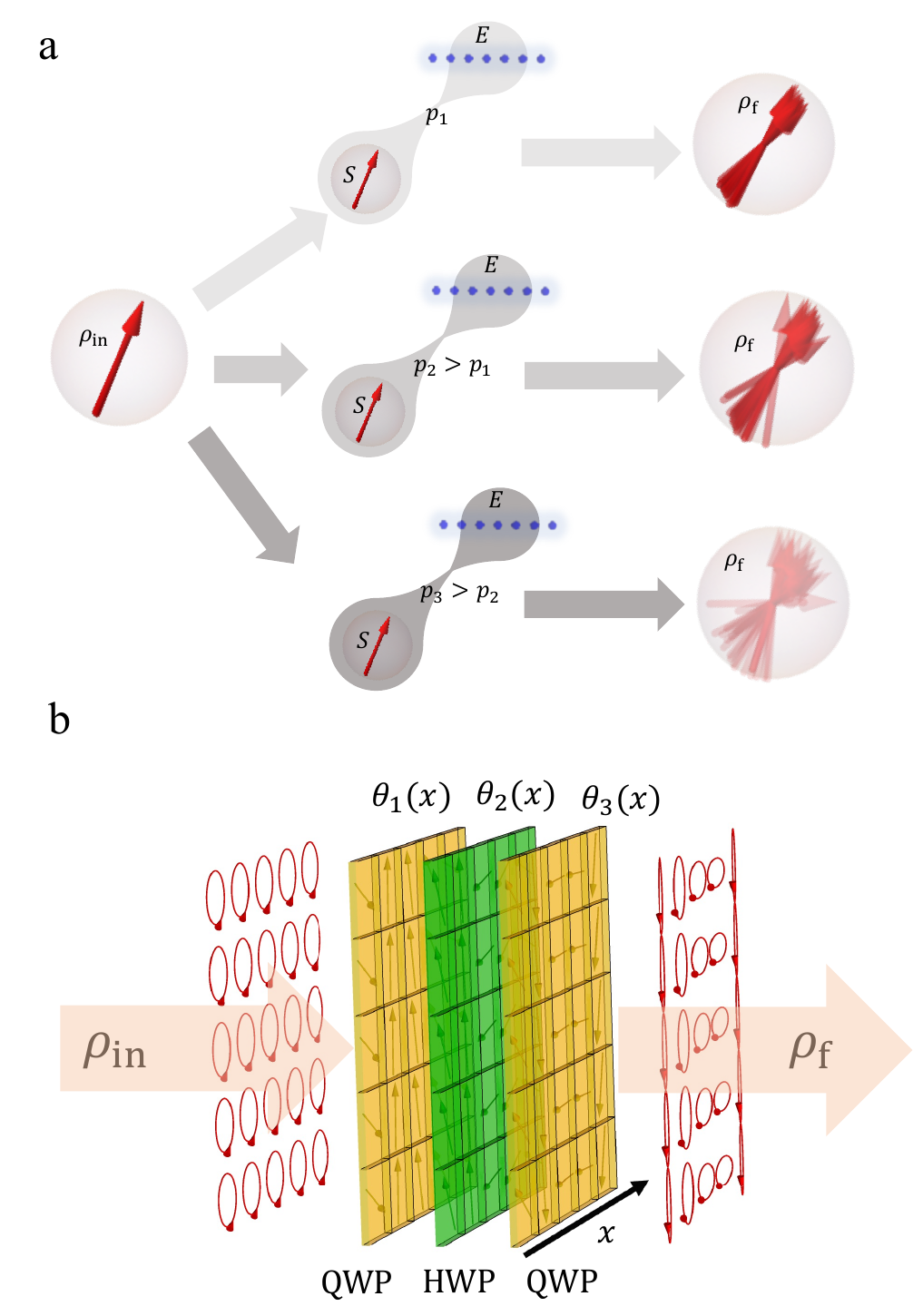}
    \caption{\textbf{Concept of the experiment.} \textbf{a}~A qubit initialized in a pure state $\rho_\text{in}$ interacts with an external environment ($E$) featuring discrete lattice symmetry. The interaction strength is controlled by the parameter $p$, which can be tuned to produce different output mixed states $\rho_\text{f}$. \textbf{b}~Our approach relies on the use of three liquid-crystal metasurfaces, each characterized by an optic-axis pattern ${\theta_i(x)}$, where $x$ encodes the position reciprocal coordinate via the mapping $x=\Lambda q/(2\pi)$. In this implementation, the qubit is encoded into light polarization, and the metasurfaces implement the system-environment interaction as a polarization-position coupling.}
    \label{fig:theory}
\end{figure}

The dynamics of a quantum system (here, a single qubit) is most generally described by a quantum channel ${\mathcal N}$, describing the noisy evolution of the system, that maps an input state $\rho_\text{in}$ into the final state ${\rho_\text{f} = \mathcal{N}(\rho_\text{in})}$. Mathematically, ${\mathcal N}$ corresponds to a completely positive, trace-preserving map, which can be specified by a set of Kraus operators ${\{A_k\}}$ such that
\begin{equation}\label{kraus}
    \rho_\text{f} = \sum_{k} A_k \rho_\text{in} A_k^\dag,
\end{equation}
under the trace-preserving condition ${\sum_k A_k A_k^{\dagger}=\sigma_0}$, where $\sigma_0$ is the ${2\times2}$ identity operator~\cite{nielsen2010quantum}. 
An equivalent description is to consider the qubit as a system $S$ interacting with an environment $E$. The qubit and environment define a closed system ${\Omega=S+E}$, described by a pure state undergoing a unitary evolution (see Fig.~\ref{fig:theory}\textbf{a}). This corresponds to choosing a \emph{purification} of the system ${S}$. The choice of purification (or, equivalently, of the environment $E$) is not unique~\cite{nielsen2010quantum}. We will show in the following that it is always possible to model the environment as an infinite-dimensional Hilbert space $\mathcal{H}_{E}$ spanned by a discrete set of states $\{\ket{m}\}$, with ${m\in 	\mathbb{Z}}$. These states can be interpreted as the eigenstates of the position operator on an infinite 1D lattice. This suggests that the system-environment interaction occurs in the form of a translation-invariant unitary operator $U$. Accordingly, the latter can be block-diagonalized in the reciprocal space: 
\begin{equation}
    U = \int^{2\pi}_{0} \text{d}q\;\mathcal{U}(q)\ketbra{q}{q} ,
    \label{eq:Un}
\end{equation}
where ${\ket{q}:=\sum_m \exp(i q m)\ket{m}/\sqrt{2\pi}}$ is a quasi-momentum eigenstate and ${\mathcal{U}(q)}$ is an SU(2) operator that acts on the Hilbert space of the qubit $\mathcal{H}_S$. 

Without loss of generality, we can assume that the environment is initialized in a pure state ${\rho_{E,\text{in}}=\ketbra{m=0}{m=0}}$, so that the global initial state is ${\rho_{\Omega,\text{in}}=\rho_{S,\text{in}}\otimes \rho_{E,\text{in}}}$. Applying $U$ to the input state, we obtain the following:
\begin{eqnarray}
    \rho_{\Omega,\text{f}}&=&U \rho_{\Omega,\text{in}} U^\dag \cr
    &=&\frac{1}{2\pi}\iint\text{d}q\,\text{d}q' \;\mathcal{U}(q)\rho_{S,\text{in}}\,\mathcal{U}^\dag(q')\ketbra{q}{q'},
\end{eqnarray}
where the integrals are taken over the interval ${[0,2\pi]}$. After the interaction, the state of the qubit can be retrieved by taking the partial trace of the evolved density operator with respect to the environment in the quasi-momentum representation:
\begin{eqnarray}
    \label{unitary}
    \rho_\text{f}=\text{tr}_E(\rho_{\Omega,\text{f}})&=&\int^{2\pi}_{0} \,\text{d}q\,\bra{q}\rho_{\Omega,\text{f}}\ket{q}\cr
    &=&\frac{1}{2\pi} \int^{2\pi}_{0} \text{d}q\,\mathcal{U}(q)\rho_\text{in}\, \mathcal{U}^\dag(q),
\end{eqnarray}
where the system label $S$ has been omitted. Our goal is to design $\mathcal{U}(q)$ such that the evolved density operator $\rho_\text{f}$ of the qubit is equivalent to the outcome of a target noise operation $\mathcal{N}$. Mathematically, this corresponds to solving
\begin{equation}
\frac{1}{2\pi} \int^{2\pi}_{0} \text{d}q\,\mathcal{U}(q)\rho_\text{in}\, \mathcal{U}^\dag(q)=\sum_kA_k\rho_\text{in}A_k^\dag,
\label{eqn:maineq1}
\end{equation}
under the constraint that $\mathcal{U}(q)$ is an SU(2) matrix at each $q$. 

To this aim, we decompose ${\mathcal{U}(q)}$ and each Kraus operator $A_k$ into linear combinations of the identity and Pauli matrices ($\sigma_1,\sigma_2,\sigma_3$), a set of complete basis for the $2\times 2$ matrices: 
\begin{eqnarray}
\mathcal{U}(q)&=&u_0(q)\sigma_0-i(u_1(q)\sigma_1+u_2(q)\sigma_2+u_3(q)\sigma_3),\cr
A_k&=&c_0^{(k)}\sigma_0-i(c_1^{(k)}\sigma_1+c_2^{(k)}\sigma_2+c_3^{(k)}\sigma_3),
\label{eqn:decomposition}
\end{eqnarray}
which yields (see Eq.~\eqref{eqn:maineq1}),
\begin{equation}
    \sum_{i,j} \frac{1}{2\pi}\int^{2\pi}_{0} \text{d}q\, u_i(q) u_j(q) \sigma_i \sigma_j = \sum_{i,j}\sum_k c_i^{(k)}c_j^{*(k)} \sigma_i\sigma_j,\qquad
    \label{eqn:eqpauli}
\end{equation}
where $*$ denotes complex conjugation. Equation~\eqref{eqn:eqpauli} corresponds to a system of 16 equations of the following form:

\begin{equation}
  \frac{1}{2\pi}  \int^{2\pi}_{0} \text{d}q\, u_i(q) u_j(q) = \sum_{k} c_i^{(k)}c_j^{*(k)},
\label{eqn:system}
\end{equation}
for ${ \lbrace{i,j\rbrace}=0,1,2,3}$, that can be solved numerically for the coefficients $\{u_i(q)\}$ 
under the unitary constraint
\begin{equation}
  u_0^2(q)+u_1^2(q)+u_2^2(q)+u_3^2(q) = 1,
\end{equation}
for all $q\in [0,2\pi]$ (see Methods for details). Importantly, the solution holds for any qubit initialization; therefore, we engineer an effective simulation of the target noise.

Once the global unitary has been identified, an experimental setting can be created to implement such an operator. In our photonic setup, the qubit is encoded into light polarization, while the environment state $\ket{m}$ corresponds to an optical mode carrying $m$ units of transverse momentum ${\Delta k_\perp=2\pi/\Lambda}$, where $\Lambda$ is the spatial period of the beam transverse profile. This corresponds to mapping the quasi-momentum coordinate into the photon transverse position: ${q=2\pi x/\Lambda}$~\cite{DErrico2020}. With this encoding scheme, arbitrary unitary operators in the form of Eq.~\eqref{eq:Un} can be simulated using a minimal stack of three patterned waveplates (see Fig.~\ref{fig:theory}\textbf{b}). This technique has been introduced in Ref.~\cite{DiColandrea2023} as a compact solution to simulate ultra-long quantum walks. It is well-known that a sequence of a half-wave plate (HWP) sandwiched between two quarter-wave plates (QWPs) can implement an arbitrary polarization transformation by adjusting the orientation of the optic axes ${\theta_i}$ of each waveplate~\cite{Simon1990}. In the circular polarization basis, where ${\ket{L}:=(1,0)^T}$ and ${\ket{R}:=(0,1)^T}$ are the left and right circular polarization states, respectively, and $T$ stands for transpose, the general transformation matrix associated with a waveplate can be written as
\begin{eqnarray}
    W_i(\theta_i)&=&\cos\left(\frac{\delta_i}{2}\right)\sigma_0\cr
    &+&i\sin\left(\frac{\delta_i}{2}\right)(\cos(2\theta_i)\sigma_1 +\sin(2\theta_i)\sigma_2),
\label{eq:waveplate}
\end{eqnarray}
where the subscript ${i=1,2,3}$ labels individual waveplates, ${\delta_1=\delta_3=\pi/2}$ are the optical retardations of the QWPs $W_1$ and $W_3$, ${\delta_2=\pi}$ is the retardation of the HWP $W_2$, and $\theta_i$ are the optic-axis orientation angles. The required patterns $\theta_i(x)$ can be determined by solving ${W_3W_2W_1=\mathcal{U}}$ at each $q$ (see Methods). 
Patterned waveplates are fabricated as liquid-crystal metasurfaces, tunable nematic liquid-crystal samples, with the optic axis locally patterned via a well-established photoalignment technique~\cite{Rubano2019}. 

\begin{figure}
    \centering
    \includegraphics[width=\columnwidth]{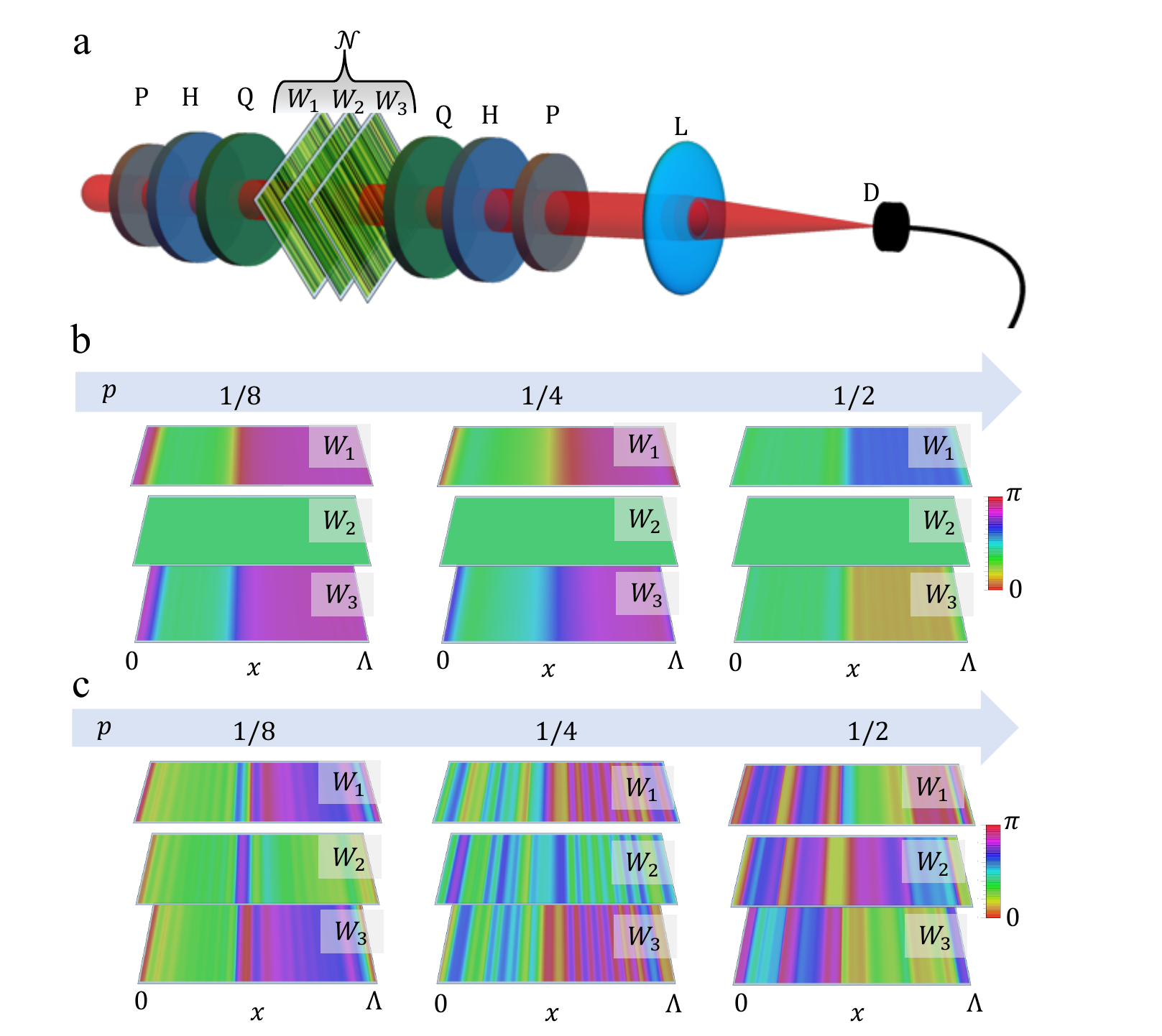}
    \caption{\textbf{Experimental setup.} \textbf{a}~Sketch of the experimental setup. P:~polarizer, H:~half-wave plate, Q:~quarter-wave plate, L:~lens, D:~detector. The noise $\mathcal{N}$ is implemented by a sequence of three liquid-crystal metasurfaces $W_1$--$W_2$--$W_3$. In our experiment, we set ${\Lambda=2.5}$~mm. The input beam waist was ${w_0\simeq \Lambda}$, corresponding to a pure localized environment. \textbf{b}~Optic-axis patterns of the plates used for phase flip, bit flip, and bit-phase flip, for different values of the coupling parameter $p$. \textbf{c}~Patterns for the depolarization channel. The hue color coding indicates the orientation $\theta(x)$ of the liquid-crystal optic axis on the metasurfaces plane.}
    \label{fig:setup}
\end{figure}

\begin{figure*}
    \centering
    \includegraphics[width=\textwidth]{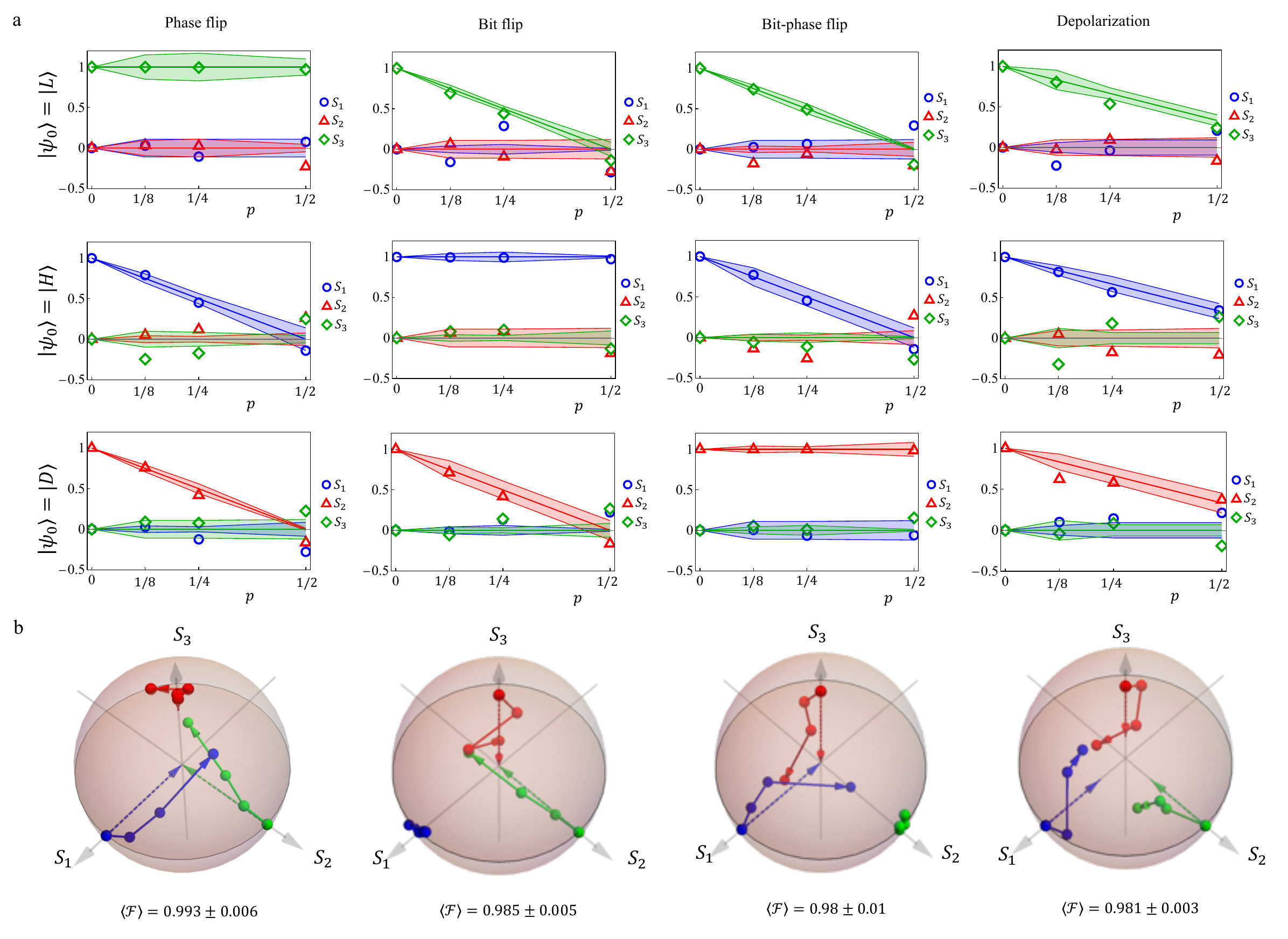}
    \caption{\textbf{Experimental results.} \textbf{a}~Output qubit states in terms of the measured Stokes parameters $S_1,\,S_2,\, S_3$ (empty markers), with varying coupling strength $p$. The ${p=0}$ case is included as a guide and corresponds to the nominal input state. Continuous lines show the theoretical trends, with error bands representing confidence intervals accounting for experimental imperfections. The latter have been evaluated with Monte Carlo simulations, where the relative alignment of the plates and their optical retardation have been considered as Gaussian random variables with a 5\% standard deviation. The bands are the standard deviations on the Stokes parameters extracted from 100 realizations of simulated experiments. Results are shown for the processes of phase flip, bit flip, bit-phase flip, and depolarization, with three different pure input states, ${\ket{H}}$, ${\ket{D}}$, and ${\ket{L}}$, respectively oriented along the $S_1$, $S_2$, and $S_3$ axis of the Bloch sphere. \textbf{b}~Qubit trajectories (as function of $p$) on the Bloch sphere. Dashed arrows are the theoretical predictions, while joined spheres are the experimental results. Different colors correspond to different input states:  ${\ket{H}}$ (blue), ${\ket{D}}$ (green), and ${\ket{L}}$ (red). In the insets, we report the average fidelities between the reconstructed and expected states.  
    }
    \label{fig:exp}
\end{figure*}

We demonstrate our approach by fabricating stacks of plates simulating different instances of quantum noise, specifically, phase flip, bit flip, bit-phase flip, and depolarization channels. These are defined by the following sets of Kraus operators. For the first three channels, ${A_0=\sqrt{1-p}}\,\sigma_0$ and ${A_1=\sqrt{p}\,\sigma_i}$, with ${i=1,2,3}$ for bit flip, bit-phase flip, and phase flip, respectively. The depolarization channel is instead characterized by the Kraus operators $\{A_0,A_1,A_2,A_3\}$, where ${A_0=\sqrt{1-p}\,\sigma_0}$ and ${A_i=\sqrt{p/3}\,\sigma_i}$~\cite{nielsen2010quantum}. Each channel has been implemented for three values of the system-environment coupling parameter, ${p=1/8,\,1/4,\,1/2}$, corresponding to the weak, intermediate, and strong interaction regimes, respectively. 

The experimental setup is outlined in Fig.~\ref{fig:setup}\textbf{a}. A 810~nm laser beam crosses a state-preparation stage consisting of a polarizer (P), a half-wave plate (H), and a quarter-wave plate (Q). For each noise realization, we choose as input states $\ket{L}$, $\ket{H}$, and $\ket{D}$, where ${\ket{H}=(\ket{L}+\ket{R})/\sqrt{2}}$ and ${\ket{D}=(\ket{L}+i\ket{R})/\sqrt{2}}$ are the horizontal and diagonal polarization states, respectively. The noise $\mathcal{N}$ is implemented by three liquid-crystal metasurfaces $W_1$--$W_2$--$W_3$. The output qubit-polarization state is analyzed with a sequence Q--H--P, measuring the transmitted intensity with a power-meter detector (D). This allows us to retrieve the three Stokes parameters providing the representation of the qubit state on the Bloch sphere~\cite{James2001}. The optic-axis patterns of the metasurfaces used to simulate phase flip, bit flip, and bit-phase flip are plotted in Fig.~\ref{fig:setup}\textbf{b}. These channels feature the same Kraus operators decomposition up to a basis rotation that can be implemented with uniform waveplates, and can therefore be simulated with the same sets of plates. The patterns for the depolarization channels are provided in Fig.~\ref{fig:setup}\textbf{c}. It is evident that the complexity of the patterns increases with the complexity of the simulated interaction. 

The experimental results are illustrated in Fig.~\ref{fig:exp}, where we plot the measured Stokes parameters (see Fig.~\ref{fig:exp}\textbf{a}) and the corresponding trajectories (see Fig.~\ref{fig:exp}\textbf{b}) of the qubit states on the Bloch sphere as a function of the noise level $p$. Deviations from the theory are systematic errors due to imperfect alignment of the plates and fabrication defects, in agreement with Monte Carlo simulations whose results are reported as error bands around the theoretical lines in Fig.~\ref{fig:exp}\textbf{a}. Overall, the average fidelity $\mathcal{F}=\Tr\left(\sqrt{\sqrt{\rho_\text{th}}\rho_\text{exp}\sqrt{\rho_\text{th}}}\right)^2$ between the measured states ${\rho_\text{exp}}$ and the expected ones ${\rho_\text{th}}$ is always above 98\%, proving that our devices reproduce the target channels with high accuracy. The results for phase flip, bit flip, and bit-phase flip show that the action of the noise keeps unaltered the qubit states initialized in the eigenstate of $A_1$, while it pushes other inputs toward the origin of the Bloch sphere. The depolarization channel displays a similar behavior for all pure input states~\cite{nielsen2010quantum}. 

\section{Conclusions}
In this work, we demonstrated a general approach for engineering arbitrary noise operations on a qubit system, based on suitably coupling the qubit with a lattice-like environment.
We provided a systematic procedure for constructing a global unitary operator corresponding to the desired process on the qubit state. In our photonic demonstration, light polarization played the role of a qubit, and the environment was encoded into a set of transverse modes, with the interaction occurring in the form of a space-dependent polarization transformation, engineered via liquid-crystal metasurfaces. The patterns of the metasurfaces have been inverse-designed to simulate typical noise operations in different interaction regimes.

Our findings pave the way for innovative solutions for studying open quantum systems. Potential applications involve the simulations of quantum machines, such as quantum refrigerators, protocols of quantum thermodynamics, chiral dynamics in open systems, and geometric-phase and topological effects in non-Hermitian evolutions. Furthermore, the controlled simulation of common forms of noise could foster the development of effective strategies for quantum error correction~\cite{Campbell2017}. Another interesting prospect concerns the generalization of our scheme to a larger number of qubits, where the emerging concepts of local passivity~\cite{PhysRevE.90.012127} and ergotropy~\cite{PhysRevA.107.012405,PhysRevLett.133.150402} could be experimentally observed. Nevertheless, one could use the same setup to implement a specific noise on a single qubit and test its effects on entangled states.
\section*{Methods}
\subsection{Numerical optimization}
To solve the system of equations given by Eq.~\eqref{eqn:system}, we expand each coefficient ${\{u_i(q)\}}$ in the Fourier basis:
\begin{equation}
u_i(q)=u_i^{(0)}+\sum_{n=1}^{N}\left( u_{ic}^{(n)}\cos\left( nq \right) + u_{is}^{(n)}\sin\left( nq \right) \right),
\label{eqn:fourier}
\end{equation}
where $N$ is the maximum number of spatial frequencies, which can be tuned. By substituting the decomposition of Eq.~\eqref{eqn:fourier} into Eq.~\eqref{eqn:system}, and performing the corresponding integrals, we obtain a simple relation for the Fourier coefficients: 
\begin{equation}
2u_i^{(0)}u_j^{(0)}+\sum_{n=1}^{N} \left(u_{ic}^{(n)}u_{jc}^{(n)}+u_{is}^{(n)}u_{js}^{(n)}\right)=2\sum_k c_i^{(k)}c_j^{*(k)},
\label{eqn:opt1}
\end{equation}
for ${\{i,j\}=0,1,2,3}$.
The system of equations given by Eq.~\eqref{eqn:opt1} can be solved numerically, under the unitary constraint ${u_0^2(q)+u_1^2(q)+u_2^2(q)+u_3^2(q)=1}$ at each quasi-momentum value. To accomplish this, an iterative solution is adopted. First, we discretize the interval $[0,2\pi]$ in ${Q=125}$ points. This value is chosen to ensure a dense sampling of the spatial period. Then, we numerically minimize the following cost function for the Fourier coefficients:
\begin{equation}
\begin{split}
f_1=\sum_{i,j}\biggl[ 2u_i^{(0)}u_j^{(0)}&+\sum_{n=1}^{N} \left(u_{ic}^{(n)}u_{jc}^{(n)}+u_{is}^{(n)}u_{js}^{(n)}\right)\\&-2\sum_k c_i^{(k)}c_j^{*(k)} \biggr]^2.
\end{split}
\end{equation}
The solutions of the first minimization routine are used as input guesses for the minimization of a second cost function, expressing the unitary constraint at each (discretized) quasi-momentum value:
\begin{equation}
f_2=\sum_{i=1}^Q \bigl(u_0^2(q_i)+u_1^2(q_i)+u_2^2(q_i)+u_3^2(q_i)-1 \bigr)^2.
\end{equation}
The second minimization output feeds the previous minimization routine for the cost function $f_1$, and so forth, until the total difference between the optimized coefficients from the two routines is less than $10^{-12}$ or when the number of iterations reaches $T=200$, whichever occurs first. This systematically allowed us to minimize both cost functions to values of order ${\approx 10^{-12}}$, by enforcing only ${N=20}$ frequencies. As detailed in the next section, the maximum number of spatial frequencies determines the complexity of the plates to be fabricated, therefore it is desirable to achieve a solution with a reasonably small number of frequencies. 

\subsection{Extracting the metasurfaces' patterns}
The nested minimization routines described above output an optimal set of Fourier coefficients for a given noise, which in turn determines the space-dependent unitary operator $U$ in the form of Eq.~\eqref{eq:Un}. After the identification ${q\leftrightarrow x}$, at each transverse position the operator $U$ implements a polarization rotation ${\mathcal{U}(x)}$ of an angle $\chi$ around the axis ${\textbf{n}=(n_1,n_2,n_3)}$:
\begin{equation}
\mathcal{U}(x)=\cos\left({\frac{\chi(x)}{2}}\right)\sigma_0-i\sin\left({\frac{\chi(x)}{2}}\right)\,\textbf{n}(x)\cdot\bm\sigma, 
\end{equation}
where ${\bm{\sigma}=\{\sigma_1,\sigma_2,\sigma_3 \}}$, and
\begin{equation}
\begin{split}
\chi&=2\arccos{u_0};\\
n_i&=\dfrac{u_i}{\sin\left(\dfrac{\chi}{2}\right)},
\end{split}
\end{equation}
for ${i=1,2,3}$ (cf.~Eq.~\eqref{eqn:decomposition}).

The optical sequence of three liquid-crystal metasurfaces ${\mathcal{L}=W_3(\theta_3)W_2(\theta_2)W_1(\theta_1)}$ can be analogously decomposed as
\begin{equation}
    \mathcal{L}(x) = \ell_0(x)\sigma_0 - i\left(\ell_1(x)\sigma_1 + \ell_2(x)\sigma_2 + \ell_3(x)\sigma_3\right),
\end{equation}
where 
\begin{equation}
\begin{split}
\ell_0 &= -\cos\alpha\cos\beta,\\
\ell_1 &= -\sin\beta\sin\gamma,\\
\ell_2 &= \sin\beta\cos\gamma,\\
\ell_3 &= \sin\alpha\cos\beta,
\end{split}
\end{equation}
with ${\alpha=\theta_1-\theta_3}$, ${\beta=\theta_1-2\theta_2+\theta_3}$, and ${\gamma=\theta_1+\theta_3}$. The dependence on the transverse coordinate $x$ is omitted for ease of notation. Imposing ${\mathcal{U}=\mathcal{L}}$ at each transverse position yields
\begin{subequations}
\begin{equation}
\cos\alpha\cos\beta=-\cos\frac{\chi}{2},
\label{eqn:eqn1}
\end{equation}
\begin{equation}
\sin\beta\sin\gamma=-\sin\frac{\chi}{2}\sin\vartheta\cos\varphi,
\label{eqn:eqn2}
\end{equation}
\begin{equation}
\sin\beta\cos\gamma=\sin\frac{\chi}{2}\sin\vartheta\sin\varphi,
\label{eqn:eqn3}
\end{equation}
\begin{equation}
\sin\alpha\cos\beta=\sin\frac{\chi}{2}\cos\vartheta,
\label{eqn:eqn4}
\end{equation}
\end{subequations}
where we used the spherical parametrization of the vector $\textbf{n}$: ${n_1=\sin\vartheta\cos\varphi}$, ${n_2=\sin\vartheta\sin\varphi}$, and ${n_3=\cos\vartheta}$.

From Eq.~\eqref{eqn:eqn2} and Eq.~\eqref{eqn:eqn3} it follows:
\begin{equation}
\gamma=\varphi-\frac{\pi}{2}.
\end{equation}
Two sets of solutions are found from Eq.~\eqref{eqn:eqn1} and Eq.~\eqref{eqn:eqn4}:
\begin{equation}
\begin{cases}
\alpha_1=\text{atan2}\left(-\sin\frac{\chi}{2}\cos\vartheta,\cos\frac{\chi}{2}\right)\\
\beta_1=\text{atan2}\left(\sin\frac{\chi}{2}\sin\vartheta,-\sqrt{1-\sin^2\frac{\chi}{2}\sin^2\vartheta}\right)\\
\gamma_1=\varphi-\pi/2
\end{cases}
\end{equation}

\begin{equation}
\begin{cases}
\alpha_2=\pi+\alpha_1\\
\beta_2=\pi-\beta_1\\
\gamma_2=\gamma_1
\end{cases}
\end{equation}
where atan2(${x,y}$) is the two-argument arctangent function, which distinguishes between diametrically opposite directions. From the expressions for $\alpha$, $\beta$, and $\gamma$, the modulations for the metasurfaces' optic-axis patterns $\theta_1$, $\theta_2$, and $\theta_3$ can be finally extracted.
A modulation obtained from a single set of solutions may exhibit discontinuities. To avoid this issue, a dedicated routine automatically switches the solution to be picked whenever the values of the angles feature a sudden jump~\cite{DiColandrea2023}. 

\newpage
\noindent \textbf{Acknowledgments.}
This work was supported by Canada Research Chairs (CRC) and Quantum Enhanced Sensing and Imaging (QuEnSI) Alliance Consortia Quantum grant. FDC further acknowledges support from the PNRR MUR project PE0000023-NQSTI.
\\
\newline \noindent \textbf{Disclosures.}
The authors declare no conflicts of interest.
\\
\newline \noindent \textbf{Data availability.} The source code and data underlying the results presented in this paper can be obtained from the corresponding author upon reasonable request.
\\
\newline \noindent \textbf{Author Contributions.}
FDC conceived the idea and developed the theory, with contributions from ADE and MA. FDC, TJ, and JG fabricated the metasurfaces. FDC, JG, and DP performed the experiment. FDC, TJ, and JG analyzed the data. FDC and ADE prepared the first version of the manuscript. ADE and EK supervised the project. 
\newpage
\clearpage

\newpage
\clearpage
\bibliography{main}

\end{document}